\documentclass[reprint,showpacs,amsmath,amssymb,aip,jcp]{revtex4-1}

\usepackage{graphicx}
\usepackage{dcolumn}
\usepackage{bm}

\usepackage{color}

\newcommand{\bra}[1]{\langle #1|}
\newcommand{\ket}[1]{|#1\rangle}

\begin{document}


\title{Visualization and Interpretation of Attosecond Electron Dynamics in
Laser-Driven Hydrogen Molecular Ion using Bohmian Trajectories}
\thanks{
The following article has been submitted to The Journal of Chemical Physics. 
After it is published, it will be found at http://jcp.aip.org/.
}

\author{Norio Takemoto}
\email{norio@jilau1.colorado.edu}
\author{Andreas Becker}%
\affiliation{%
   JILA and Department of Physics, University of Colorado, 440 UCB, Boulder,
   CO 80309-0440
 }%

\date{\today}

\begin{abstract}
We analyze the attosecond electron dynamics in hydrogen molecular ion driven by an external 
intense laser field using ab-initio numerical simulations of the corresponding time-dependent
Schr{\"{o}}dinger equation and Bohmian trajectories. 
To this end, we employ a one-dimensional model of the molecular ion in which the 
motion of the protons is frozen. The results of the Bohmian trajectory calculations 
do agree well with those of the ab-initio simulations and clearly visualize the electron transfer 
between the two protons in the field. In particular, the Bohmian trajectory calculations confirm the 
recently predicted attosecond transient localization of the electron at one of the protons and 
the related multiple bunches of the ionization current within a half cycle of
the laser field.
Further analysis based on the quantum trajectories shows that the electron dynamics in the 
molecular ion can be understood via the phase difference accumulated between the Coulomb wells 
at the two protons. Modeling of the dynamics using a simple two-state system leads us to 
an explanation for the sometimes counter-intuitive dynamics of an electron
opposing the classical force of the electric field on the electron.
\end{abstract}

\maketitle

\section{Introduction}

The causal interpretation of quantum mechanics by de Broglie and Bohm
provides the concept of trajectories for the dynamics of microscopic
objects~\cite{physRev85_166,physRev85_180}.
These trajectories, called Bohmian trajectories or quantum trajectories,
are navigated by the wavefunction.
Conversely, if we regard the probability density (i.e., 
the squared modulus of the wavefunction) of the system 
as a fluid, the flow of this fluid can be elucidated by the quantum
trajectories~\cite{Holland1993,Wyatt2005}. 
It is this characteristics of the Bohmian trajectories that
we want to utilize in this article to visualize and analyze a recently
revealed counter-intuitive
electronic motion in H$_2^+$ molecular ion exposed to intense laser light on
an attosecond time scale~\cite{jcp110_11152,prl101_213002,prl105_203004}.

Previously, we found in ab-initio numerical simulations
that H$_2^+$ at intermediate internuclear distances (i.e., between the
equilibrium distance and the dissociation limit) in an intense infrared laser pulse
shows multiple bursts of ionization within a half-cycle of the laser field
oscillation~\cite{prl105_203004}.
This ionization dynamics contradicts the widely accepted picture
of strong-field ionization, namely that an electron leaves
the atom or molecule with largest probability at the peaks of the oscillating
electric field of the laser. For example, in the often used tunnel ionization
picture the electron tunnels through the barrier created by the binding potential
of the ionic core and the electric potential of the laser field. This tunnel
barrier is, of course, thinnest when the electric field strength is strongest, which
leads to the above mentioned expectation for the most likely time instant of
the electron escape.
We identified that the multiple ionization bursts are induced by a previously
reported ultrafast transient localization of the electron
density at one of the protons~\cite{jcp110_11152}.
This attosecond dynamics can cause that the
electron density near the tunnel barrier in the molecular ion is highest
when the external field strength is below its peak strength. Correspondingly,
the electron does not tunnel most likely at the maxima of the field but at
other time instants.

We confirmed and extended earlier interpretations~\cite{pra52_r2511,jcp110_11152}
that the internal dynamics
of the electron is a result of a strong and exclusive trapping of the population
within a pair of states of opposite parity, so-called charge resonant 
states~\cite{jcp7_20}.
Consequently, the attosecond localization dynamics of the electron in the hydrogen
molecular ion driven by the laser field can be successfully reproduced using a
two-state model~\cite{prl105_203004}. 
The dynamics can be also understood by the phase-space flow of the electron probability
density regulated through so-called momentum gates which are shifted in time
by the vector potential of the external laser field~\cite{prl101_213002}.

However, results of numerical simulations for the flow of the
electron probability density often do not reveal many details and cannot provide
much further insights into the internal electron dynamics in the molecular ion.
We therefore use the concept of Bohmian trajectories 
to provide a complementary picture
of the dynamics. By analyzing the motion
of the trajectories we furthermore clarify that
it is the phase difference of the wavefunction at the two protons
that is the origin of the force which is sometimes driving the electron
in the direction opposite to the strong electric field of the laser light.

The rest of the paper is organized as follows. In Sec.~\ref{sec: Theory},
we present the model for H$_2^+$ used for our analysis.
In Sec.~\ref{sec: Results and Discussion}, we present the results for the Bohmian
trajectories and compare them with those for the electron probability densities obtained 
from ab-initio numerical simulations. We identify the phenomena of transient 
electron localization and multiple ionization bursts in the time evolution of the trajectories.
In Sec.~\ref{sec: Analysis} we make use of the Bohmian trajectory calculations to provide 
an analysis of the sometimes counter-intuitive electron dynamics in the hydrogen molecular ion. 
Finally, Section~\ref{sec: Conclusion} concludes the paper.

\section{Theory}\label{sec: Theory}

In recent studies of laser induced dynamics of the hydrogen molecular ion
the full three-dimensional electronic motion \cite{pra80_023426} or the
two-dimensional electronic motion in cylindrical coordinates along with
nuclear motion in one dimension have been taken into
account \cite{pra52_2977,prl93_163601,prl99_083002}. In the present study,
we consider a simpler one-dimensional (1D) model of H$_2^+$ in which the
positions of the protons are fixed in space. We have shown that the electron
localization dynamics inside the molecular ion as well as the phenomenon of
multiple ionization bursts does not change in higher dimensional models
\cite{prl105_203004,takemoto_tobepublished}.
In particular, we found that the nonadiabatic coupling between the electronic and nuclear
motions is not essential for the internal electron dynamics.

\subsection{1D fixed-nuclei model of H$_2^+$}

In our 1D model of H$_2^+$ with fixed positions of the protons,
the internuclear axis
was assumed to be parallel to the polarization direction of the linearly
polarized laser light. The Hamiltonian of this model system is given by
(Hartree atomic units, $e=m=\hbar=1$,
are used throughout this article unless noted otherwise)
\begin{equation}\label{eq: Hamiltonian}
H(t) =
-\frac{1}{2} \frac{\partial^2}{\partial z^2}
+ V_{\text{C}}(z;R) + V_{\text{L}}(z,t),
\end{equation}
where $z$ is the electron position measured from the center-of-mass of the two
protons.
The Coulomb interaction between the electron and the two protons was
approximated by the soft-core
potential~\cite{physRevA38_3430,physRevA53_2562},
\begin{equation}
V_{\text{C}} = -\frac{1}{\sqrt{ (z+R/2)^2 + a}} -\frac{1}{\sqrt{ (z-R/2)^2 + a}},
\end{equation}
where $a$ is the soft-core parameter.
The laser-electron interaction was expressed in the length gauge as
\begin{equation}
V_{\text{L}}(z,t)  = zE(t).
\end{equation}
The laser electric field $E(t)$ is 
related to the vector potential $A(t)$ by
\begin{equation}\label{eq: E=-dAdt}
\begin{split}
E(t) =& - \frac{\partial A(t)}{\partial t}  \\
      =& -\left[ \frac{\partial f_A(t)}{\partial t} \sin(\omega t+\varphi)
                     +f_A(t) \omega \cos(\omega t+\varphi)
               \right] ,
\end{split}
\end{equation}
where we used the following form of the vector potential:
\begin{align}
A(t) =&  f_A(t) \sin\left[\omega \left(t-\frac{T}{2}\right) + \varphi \right],
\label{eq: vecpot}
\\
f_A(t) =&
\begin{cases}
A_0 \sin^2( \frac{\pi t}{T}) & (0 \leq t \leq T) \\
0                          & (\text{otherwise})
\end{cases}
\label{eq: vecpot envelope}
.
\end{align}
The full-width at half-maximum (FWHM) of this envelope function, $f_A(t)$,
is equal to $T/2$.


\subsection{Propagation of the wavefunction and the quantum trajectories}

With the Hamiltonian given as above,
the wavefunction $\Psi(z,t)$ was propagated according to
the corresponding time-dependent Schr\"{o}dinger equation (TDSE),
\begin{equation}\label{eq: TDSE 1D model}
i\frac{\partial}{\partial t} \Psi(z,t) = H(t) \Psi(z,t).
\end{equation}
This TDSE 
was solved numerically using the second-order split-operator
method on the Fourier grid~\cite{jcp91_3571,jcp99_1185,jcp99_8680}.
The spatial and temporal grid intervals used for the simulations
were $\Delta z=0.152$ and $\Delta t = 0.0245$ or smaller.

At the same time as the wavefunction $\Psi(z,t)$
was propagated in time,
the Bohmian trajectories $\{z_j(t)| j=1,\dots,N_{\text{traj}}\}$ 
were propagated as well by solving the equation of motion,
\begin{equation}\label{eq: EOM qtraj}
\frac{d z_j}{dt} = v(z_j(t), t), 
\end{equation}
where the velocity field $v(z,t)$ is given by the phase gradient of the
wavefunction,
\begin{equation}
\Psi(z,t) = R(z,t) \exp(i S(z,t))
\end{equation}
with $R\geq 0$ and $S \in \mathbb{R}$,
as
\begin{equation}\label{eq: velocity field}
v(z,t)
= \frac{\partial S}{\partial z}.
\end{equation}
The following identity was utilized in the actual computation:
\begin{equation}
  \frac{\partial S}{\partial z}
= \text{Im}\left[\frac{1}{\Psi} \frac{\partial \Psi}{\partial z}\right].
\end{equation}
The ordinary differential equation (\ref{eq: EOM qtraj})
with respect to $t$ was solved
by the fourth order Runge-Kutta scheme~\cite{Press_etal1992} 
with the fixed step size of $2\Delta t$, i.e., twice the 
time step of the wavefunction propagation.
The wavefunction value at every other step of its propagation 
was used to evaluate the velocities of the trajectories
at the mid-point of one Runge-Kutta step
to achieve the fourth order accuracy.
We may note parenthetically that expression (\ref{eq: velocity field}) 
for the velocity field is valid for the wavefunction in the length gauge
representation used in the present study. 
In the velocity gauge representation, the velocity field
is given by $v(z,t) = \partial S/\partial z -A(t)$~\cite{Holland1993}.

The initial positions of the quantum trajectories were distributed 
at a regular interval, $\delta z$, and for each trajectory 
we assigned the weight,
\begin{align}
w_j =& \int_{\Omega_j(t_0)} dz |\psi(z, t_0)|^2 ,
\label{eq: Bohm weight}
\\
\Omega_j(t_0) 
=& \left\{ z \left|   z_j(t_0)-\frac{\delta z}{2} < z 
               < z_j(t_0)+\frac{\delta z}{2} \right.\right\} .
\end{align}
At the limit of $\delta z \rightarrow 0$, we may consider that the 
weight $w_j$ assigned at the initial time ($t=t_0=0$) is conserved over the 
time evolution~\cite{chemPhysLett364_562,jcp118_2482}.

We note here that 
the quantum trajectories can actually be propagated without
pre-calculating the wavefunction on the entire simulation
space~\cite{prl82_5190, jcp111_2423, Wyatt2005,chemPhys370_4, pra82_021404}.
This 'synthetic' propagation technique has recently attracted much attention
in view of the development of efficient computational schemes to simulate
quantum mechanical systems with a large number of degrees of freedom.
However, in the present work, for the sake of focusing on the comparison of the
Bohmian trajectories with the results for the electron probability density
we propagated both the solution of the TDSE
as well as the trajectories.

\section{Visualization of the internal electron dynamics}
\label{sec: Results and Discussion}


In this section we compare the results from the Bohmian trajectory calculations
with electron probability densities obtained by integrating the TDSE
for the 1D H$_2^+$ model interacting with a linearly polarized intense laser pulse.
For this exemplary comparison we have chosen the distance between the two protons as
$R=7$ and considered a laser pulse with a peak intensity of $4\times 10^{13}$ W/cm$^2$, 
a wavelength of $1064$ nm, a full duration of $T=10$ cycles, and 
a carrier-to-envelope phase of $\varphi=0$. 
We set the soft-core parameter as $a=2.0$ so that the 
energies of the ground and first-excited electronic states of the 1D model
($-0.519$ and $-0.491$, including the $1/R$ nuclear repulsion)
best reproduce the exact
values\cite{atDataNucDataTables2_119} ($-0.506$ and $-0.496$)
for the actual H$_2^+$ in 3D space at $R=7$.
The initial positions of $100$ quantum trajectories were distributed over
$-10 \leq x \leq 10$ at a regular interval $\delta z=0.202$, and their weights $\{w_j\}$ 
were determined via Eq.~(\ref{eq: Bohm weight}).

\begin{figure}[tbhp]
\begin{center}
\end{center}
\includegraphics[width=.9\linewidth]{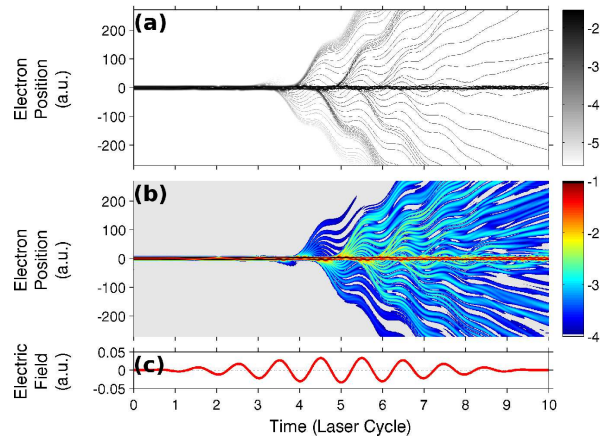}
\caption{Comparison of the quantum trajectories (a) and the 
electron probability density (b) for the 1D model of H$_2^+$ at $R=7$
for the interaction of H$_2^+$ with an electric field (c) of a laser pulse 
(peak intensity $4\times 10^{13}$ W/cm$^2$, wavelength $1064$ nm,
duration $T=10$ cycles, and CEP $\varphi=0$).
In panel (a), the gray-scale color of each trajectory indicates
$\log_{10}w_j$. In panel (b), the color code indicates
$\log_{10}|\Psi|^2$, where $\Psi$ is obtained by solving 
the TDSE (\ref{eq: TDSE 1D model}).
}
\label{fig: density_qtrajs}
\end{figure}

In Fig.~\ref{fig: density_qtrajs} 
the Bohmian trajectories (Fig.~\ref{fig: density_qtrajs}(a)) 
are presented along with the electron 
probability density (Fig.~\ref{fig: density_qtrajs} (b)) 
as a function of time. For the sake of comparison,
the electric field of the laser pulse is shown 
in Fig.~\ref{fig: density_qtrajs}(c) as well.
Subject to the intense electric field, the trajectories leave the 
core region (at $z\approx 0$) of the molecular ion
in alternating directions ($z \rightarrow \pm \infty$) 
at every half cycle of the laser field. 
It is clearly seen that 
the number of ionizing trajectories increases as the field strength increases
during the laser pulse. 
The trajectories liberated from the core region show wiggling motions
forced by the alternating electric field of the laser. Due to this 
quiver motion, some of the trajectories, depending on the time instants
of their release, are driven back to the core region and scattered off the
protons.
Please note that the results for the Bohmian trajectories visually agree very well 
with those for the electron probability density: regions of large probability density 
correspond to a large density of the Bohmian trajectories. Thus, the Bohmian 
trajectories provide a complete overall picture of the ionization process, 
including the quiver motion and the rescattering of the electron in the laser field. 
This agrees with the findings of earlier studies using Bohmian trajectories to 
describe the interaction of atoms with intense laser 
pulses~\cite{pramana51_585,euroPhysJD53_393,newJPhys11_113035,pra82_021404}.

\begin{figure}[tbhp]
\begin{center}
\end{center}
\includegraphics[width=.9\linewidth]{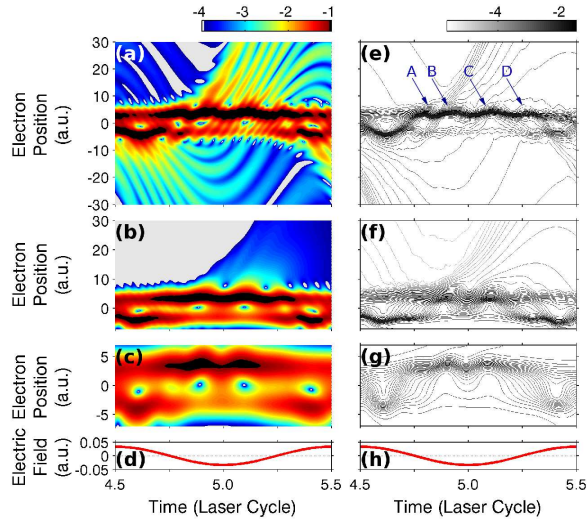}
\caption{Close-up view on the time evolution of the electron probability 
density (a-c) and the quantum trajectories (e-g) over the single laser 
cycle at the pulse peak. 
The laser electric field is shown in panels (d) and (h)
for comparison with the flow of the electron probability density and 
the quantum trajectories.
In panels (a) and (d), no wavepacket absorber was set close to the core region.
In panels (b) and (f), an absorber was set on one side of the core region 
at $-11 <z<-7$.
In panels (c) and (g), two absorbers were set on both sides of the core region
at $7 < |z| < 11$.
}
\label{fig: traj_density_c1c_gather}
\end{figure}

In Fig.~\ref{fig: traj_density_c1c_gather} we show a detailed view 
of the electron dynamics in and close to the core region 
(the protons are located at $z=\pm R/2=\pm 3.5$ a.u.) over the central field cycle 
($T/2-2\pi/\omega < t < T/2 + 2\pi/\omega$)
of the laser pulse using the electron probability density 
[Figs.~\ref{fig: traj_density_c1c_gather}(a)-\ref{fig: traj_density_c1c_gather}(c)] 
and the Bohmian trajectories 
[Figs.~\ref{fig: traj_density_c1c_gather}(e)-\ref{fig: traj_density_c1c_gather}(g)]. 
The laser electric field in the same time window is also shown in 
Figs.~\ref{fig: traj_density_c1c_gather}(d) and \ref{fig: traj_density_c1c_gather}(h). 
In the simulations for 
Figs.~\ref{fig: traj_density_c1c_gather}(a) and \ref{fig: traj_density_c1c_gather}(e),
the wavefunction masks of $\cos^{1/4}$-shape 
were placed over $270 < |z| < 300$
only to avoid the reflection of the ionized
wavepackets at the boundaries of the large simulation box,
leaving the electron dynamics in the region of $z$ shown in these panels
unaffected. Both of the electron probability density and the quantum trajectories 
show an ultrafast oscillatory motion around $5 < |z| < 10$ as indicated by, for example, 
the peaks A-D in Fig.~\ref{fig: traj_density_c1c_gather}(e).
This oscillation involves the ionization and rescattering of the electron. 
To see this we suppressed the rescattering effect on the negative $z$-side by
setting a wavefunction mask at $-11 < z < -7$
for the simulations 
in Figs.~\ref{fig: traj_density_c1c_gather}(b) and 
\ref{fig: traj_density_c1c_gather}(f)~\cite{philTransRoySocLondonA356_317,optExpress6_111,physRevA74_033405,prl105_203004}.
As a result, the peak A in Fig.~\ref{fig: traj_density_c1c_gather}(e)
disappears in Fig.~\ref{fig: traj_density_c1c_gather}(f), indicating that this
peak was created by the wavepacket driven back from $z<0$ and passing through the core
region toward $z>0$.
In Figs.~\ref{fig: traj_density_c1c_gather}(c) and \ref{fig: traj_density_c1c_gather}(g),
the wavefunction masks were set close to the core on both sides (at $7 < |z| < 11$) 
and all the rescattering wavepackets were absorbed after the initial ionization.
The two peaks B and C in Fig.~\ref{fig: traj_density_c1c_gather}(e) are still
present in Fig.~\ref{fig: traj_density_c1c_gather}(g) while peak D disappears, 
indicating that there are actually two bursts of ionization under the present
conditions. These results confirm the previously reported phenomenon of
multiple ionization bursts and attosecond electron
localization~\cite{jcp110_11152,prl105_203004}. 

As in Fig.~\ref{fig: density_qtrajs}, the quantum trajectories reflect well the
characteristics of the flow of the electron density in each pair of panels in 
Fig.~\ref{fig: traj_density_c1c_gather}. In addition, the quantum trajectories 
in Fig.~\ref{fig: traj_density_c1c_gather}(g) clearly elucidates that the 
electron probability moves back and forth between the two protons on the 
ultrafast time scale shorter than a half-cycle of the laser field.
This is not as obvious in the plot of the electron probability density
in Fig.~\ref{fig: traj_density_c1c_gather}(c). Thus, the quantum trajectories
are shown to provide further insights in the flow of the probability density.

Please note that the electronic motion 
in Fig.~\ref{fig: traj_density_c1c_gather}(g)
between the two potential wells created 
by the protons does not necessarily follow the laser-electron interaction potential 
$V_{\text{L}}$. For example, at $t=5$ cycles, 
the oscillating laser electric field $E(t)$ is
peaked in the negative direction (cf. Fig.~\ref{fig: traj_density_c1c_gather}(h)), 
and hence the slope of $V_{\text{L}}$ pushing the electron toward
$z\rightarrow +\infty$ becomes maximum. Nevertheless, we observe some bound trajectories 
propagated in the opposite direction 
from the proton located at $z=R/2=3.5$ to the other one at $z= -R/2= -3.5$ by
climbing up the potential $V_{\text{L}}$. This counter-intuitive (and classically forbidden) motion 
of the electron was noticed first in the context of coherent control of electron localization
in dissociating H$_2^+$ molecule by the Wigner
representation~\cite{prl101_213002}. 
The present results confirm this motion and visualize it using Bohmian trajectories.

\section{Analysis of the electron dynamics using Bohmian trajectories}
\label{sec: Analysis}

We have seen so far that the Bohmian trajectories clearly visualize the
transient electron localization and multiple ionization bursts within a
half-cycle of the laser field oscillation. In this section we will now 
investigate the origin of the counter-intuitive motion of the Bohmian trajectories.
To this end, we focus on the results for the intra-molecular electron transfer from 
one proton to the other, obtained by using wavepacket absorbers over $7 < |z| < 11$ 
in order to eliminate the effect of rescattering wavepackets 
(cf.\ Figs.~\ref{fig: traj_density_c1c_gather}(c) 
and \ref{fig: traj_density_c1c_gather}(g)).

\subsection{Velocity field for the Bohmian trajectories}

\begin{figure}[tbhp]
\begin{center}
\end{center}
\includegraphics[width=.9\linewidth]{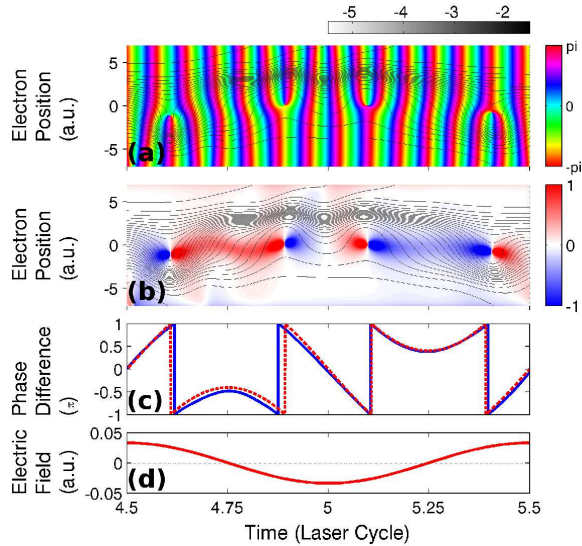}
\caption{Analysis of the intra-molecular motion of the quantum trajectories
in terms of the phase of the wavefunction $S(x,t)$ and the velocity field $v(x,t)$.
(a) Phase $S(x,t)$ and the quantum trajectories $\{z_j(t)\}$.
(b) Velocity field $v(x,t)$ and the quantum trajectories $\{z_j(t)\}$.
(c) Phase difference between the two wells calculated from the 1D model
($\alpha_{\text{LR}}(t)$, blue solid line) and from the approximate solution of the 
two-state model ($\alpha_{\text{LR}}^{(\text{2lev})}(t)$, red dashed line).
(d) Electric field of the laser pulse (wavelength $1064$ nm, peak intensity
$4\times 10^{13}$ W/cm$^2$, duration $T=10$ cycles, and CEP $\phi=0$). 
}
\label{fig: phase traj}
\end{figure}

As pointed out before, we observe that some of the trajectories 
turn their direction toward $z<0$ 
near the peak of the electric field at $t=5$ cycles 
while the classical force due to the laser electric field pushes the electron
in the positive $z$ direction. We now investigate this counter-intuitive motion in terms of
the velocity field, $v(x,t)=\partial S(x,t)/\partial x$, for the
Bohmian trajectories. Fig.~\ref{fig: phase traj}(a) shows the phase $S(x,t)$ of the 
wavefunction and Fig.~\ref{fig: phase traj}(b) the velocity field $v(x,t)$
both calculated using the same parameters of the field as before. In both Figures 
we superposed the Bohmian trajectories for further visualization.

From  Fig.~\ref{fig: phase traj}(a), we can notice that 
at a given time instant the phase of the wavefunction 
is almost constant within each of the potential wells. However, the phase propagates at different
speeds in the two wells. This causes a phase gradient around $z=0$ and, consequently,
a large absolute value of the velocity field in the region between the protons.
Over the central half ($4.75 < t < 5.25$ laser cycles) of the field cycle
shown in  Fig.~\ref{fig: phase traj},
the velocity field changes its sign at $t=4.89$, $5.0$, and $5.10$ laser cycles
and, hence, forces the Bohmian trajectories to turn their direction at these instants.
The change of sign of the 
velocity field corresponds to those instants at which the accumulated phase difference 
between the two wells equals a multiple of $\pi$. 
This interpretation is confirmed by the results for the relative phase of the wavefunction, defined as 
\begin{equation}\label{eq: alpha LR 1D numerical}
\alpha_{\text{LR}}(t) 
= \arg\left[\Psi\left(z=-\frac{R}{2},t\right)\right] 
 -\arg\left[\Psi\left(z= \frac{R}{2},t\right)\right],
\end{equation}
and shown as solid line in Fig.~\ref{fig: phase traj}(c). One can clearly see that every jump 
in the relative phase $\alpha_{\text{LR}}$ from $-\pi$ to $\pi$ corresponds to an 
abrupt sign change of $v(x,t)$.
We may note that this divergence of the $v(x,t)$ is accompanied by a node 
in the wavefunction at the same position 
[cf. Fig.~\ref{fig: traj_density_c1c_gather}(c)], and that therefore the flux
$v(x,t)|\Psi(x,t)|^2$ stays finite. 
The phase jump between $-\pi$ and $\pi$ can be interpreted as following:
If the local phase at one well gets more and more retarded (or advanced) from that at
the other well and if the absolute difference passes $\pi$, then the former should 
be now regarded as more advanced (retarded) than the latter.

\subsection{Phase difference in two-state model}

Next, we show that the phase difference $\alpha_{\text{LR}}$ 
can be approximated by a simple expression based on a two-state model.
To this end, we analyze the relative phase between the two potential wells 
in terms of the following localized 
states~\cite{physRevB47_9940,physRevA50_843,physRevB67_165301}, 
\begin{align}
\ket{\text{L}} =& \frac{1}{\sqrt{2}} \left[\ket{\text{g}}+\ket{\text{u}}\right],\\
\ket{\text{R}} =& \frac{1}{\sqrt{2}} \left[\ket{\text{g}}-\ket{\text{u}}\right],
\end{align}
where $\ket{\text{g}}$ and $\ket{\text{u}}$ are the 
ground and first-excited electronic states, respectively, of the 1D fixed-nuclei model.
Without loss of generality, we set the phases of $\ket{\text{g}}$ and $\ket{\text{u}}$ 
such that $\ket{\text{L}}$ and $\ket{\text{R}}$ are localized at $z=-R/2$ and
$z=R/2$, respectively.
By approximating the state of the system in the basis of these two
localized states as 
$\ket{\Psi(t)} 
= c_{\text{L}}(t) \ket{\text{L}} + c_{\text{R}}(t) \ket{\text{R}}$,
the time-evolution of $c_{\text{L}}(t)$ and $c_{\text{R}}(t)$
is given by
\begin{equation}\label{eq: TDSE cL cR}
i \frac{d}{dt}
\begin{pmatrix}
c_{\text{L}} \\
c_{\text{R}}
\end{pmatrix}
 = \left[ H^{(\text{2lev})}_0 + V^{(\text{2lev})}_{\text{L}} \right]
\begin{pmatrix}
c_{\text{L}} \\
c_{\text{R}}
\end{pmatrix}
,
\end{equation}
with the field-free Hamiltonian
\begin{equation}
H^{(\text{2lev})}_0 = 
-\frac{\Delta_0}{2}
\begin{pmatrix}
0 & 1\\ 
1 & 0
\end{pmatrix}
,
\end{equation}
and the interaction potential
\begin{equation}
V^{(\text{2lev})}_{\text{L}}
 = -d_{\text{gu}} E(t)
\begin{pmatrix}
1 &  0  \\
0 & -1
\end{pmatrix}
,
\end{equation}
where
$\Delta_0$ is the absolute value of the difference between 
the field-free energies of $\ket{\text{u}}$ 
and $\ket{\text{g}}$, and
$d_{\text{gu}} = -\bra{\text{g}} z \ket{\text{u}} \geq 0$
is the transition dipole moment between $\ket{\text{g}}$ and $\ket{\text{u}}$.

If the laser-molecule coupling $|d_{gu} E(t)|$ is sufficiently strong 
and/or the laser frequency $\omega$ is sufficiently larger than 
the tunnel splitting $\Delta_0$, we may approximate the solution to 
the two-state TDSE (\ref{eq: TDSE cL cR}) 
by taking $V^{(\text{2lev})}_{\text{L}}$
as the zeroth order reference Hamiltonian and omitting the $H_0$ term
\cite{prl67_516,europhysLett18_571,physRevA48_580,laserPhys3_375,JETP85_657,physRevB67_165301,jPhysB34_2371}.
Such a zeroth order solution can be easily obtained 
as~\cite{physRevB67_165301}
\begin{align}
c^{(0)}_{\text{L}}(t)
=& \exp\left\{ -i d_{\text{gu}} [A(t)-A(t_0)] \right\} c^{(0)}_{\text{L}}(t_0) 
,
\label{eq: c0L}
\\
c^{(0)}_{\text{R}}(t)
=& \exp\left\{ i d_{\text{gu}} [A(t)-A(t_0)] \right\} c^{(0)}_{\text{R}}(t_0) 
,
\label{eq: c0R}
\end{align}
where the relation of the electric field and vector potential, 
Eq. (\ref{eq: E=-dAdt}), was used. 
Then, the phase difference between the two wells can be approximated by
\begin{equation}\label{eq: def alpha 2lev}
\alpha^{(\text{2lev})}_{\text{LR}}(t) 
= \arg\left[c^{(0)}_{\text{L}}\right] - \arg\left[c^{(0)}_{\text{R}}\right].
\end{equation}
By substituting Eqs. (\ref{eq: c0L}) and (\ref{eq: c0R}) 
into Eq. (\ref{eq: def alpha 2lev}), 
and noting that $A(t_0)=0$ at the beginning of the laser pulse at $t=t_0=0$,
we obtain 
\begin{equation}\label{eq: alpha LR 2lev}
\alpha^{(\text{2lev})}_{\text{LR}}(t) 
= -2 d_{\text{gu}} A(t) 
  +\alpha^{(\text{2lev})}_{\text{LR}}(t_0).
\end{equation}
The phase difference calculated by this expression is plotted in 
Fig.~\ref{fig: phase traj}(c) as the red dashed line.
We can see that the two-state model closely reproduces 
the exact value (blue solid line) calculated numerically by
Eq. (\ref{eq: alpha LR 1D numerical}).

The condition,
$\alpha_{\text{LR}}^{(\text{2lev})}(t_{\text{turn}})= n\pi$, $n\in \mathbb{Z}$,
for the time instant $t_{\text{turn}}$ at which the velocity field 
between the two protons changes its sign
is then given by 
\begin{equation}\label{eq: tturn condition}
A(t_{\text{turn}})
= \frac{-n\pi +\alpha^{(\text{2lev})}_{\text{LR}}(t_0)}{ 2 d_{\text{gu}} }.
\end{equation}
This expression has a similar form as the condition for the time instant
$t_{\text{loc}}$ of maximum electron localization,
\begin{equation}\label{eq: tloc condition}
A(t_{\text{loc}}) = \frac{m\pi + \chi}{2d_{\text{gu}}} ,
\end{equation}
with $m\in \mathbb{Z}$
derived previously~\cite{prl105_203004} based on the series expansion of the
Floquet states for the two-state model~\cite{jPhysB34_2371}.
In fact, these two expressions are identical 
at the limit of long laser pulse duration where the 
mixing angle $\chi$ of the two Floquet states reduces to 
the initial phase difference $\alpha^{(\text{2lev})}_{\text{LR}}(t_0)$.

\subsection{Origin of the counter-intuitive electron motion}

The result of the two-state analysis in the previous subsection,
in which we consider $V^{(\text{2lev})}_{\text{L}}$ 
as the zeroth order Hamiltonian while neglecting the 
tunnel hopping term $H_0$,
provides us with an intuitive picture.
Please note that the transition dipole moment has the 
asymptotic form $d_{\text{gu}} \sim R/2$ at large $R$~\cite{jcp7_20}, 
and the difference of the diagonal elements of $V^{(\text{2lev})}$
is, hence, approximately equal to $R E(t)$, which is the difference of the 
electric potential induced by the laser field between the two wells.
Using this asymptotic form of $d_{\text{gu}}$, the phase difference
in Eq.~(\ref{eq: alpha LR 2lev}) can be rewritten as
\begin{equation}\label{eq: alpha LR 2lev by Et}
\alpha^{(\text{2lev})}_{\text{LR}}(t) 
\sim \int_{t_0}^t dt' R E(t') 
  +\alpha^{(\text{2lev})}_{\text{LR}}(t_0).
\end{equation}
This expression elucidates that the origin of the phase difference 
between the two potential wells, and hence the velocity field,
is the difference of the electric potential energies between the two
wells induced by the laser light.

\begin{figure}[tbhp]
\begin{center}
\end{center}
\includegraphics[width=.9\linewidth]{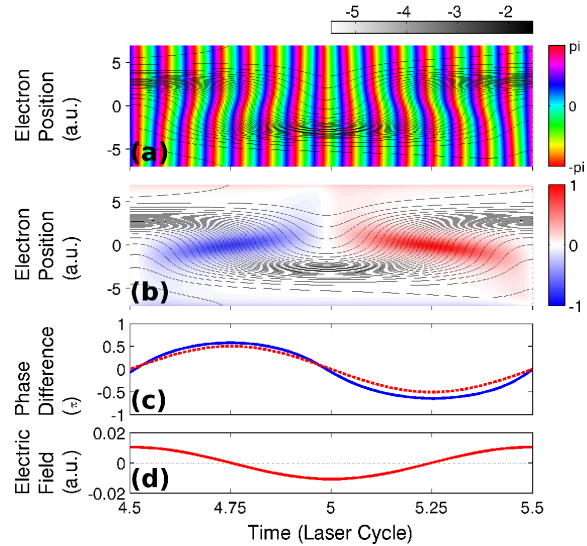}
\caption{Time evolutions of (a) the phase of the wavefunction and (b) the velocity field 
are plotted along with the quantum trajectories. 
The system was initially prepared in the ground state $\ket{\text{g}}$.
Panel (c) shows the
phase difference between the two local potential wells 
calculated from the TDSE solution (blue solid line) and
from the two-state model (red dashed line). Panel (d) shows the electric field
of the laser pulse (wavelength $1064$ nm, peak intensity $4\times 10^{12}$
W/cm$^2$, duration $T=10$ cycles, and CEP $\phi=0$).
}
\label{fig: 1locg}
\end{figure}

This interpretation of the electronic dynamics based on the zeroth order 
two-state analysis predicts that the motion of the electron should be still
counter-intuitive even at a relatively low intensity at which only one
localization per half-cycle is predicted by Eq.~(\ref{eq: tturn condition})
or (\ref{eq: tloc condition}).
This is demonstrated in Fig.~\ref{fig: 1locg}
in which the 1D model was initially prepared in the ground state,
and a laser pulse of peak intensity $4\times 10^{12}$ W/cm$^2$ was applied. 
The other parameters were the same as above. 
The initial phase difference $\alpha^{(\text{2lev})}_{\text{LR}}(t_0)$ is zero due to the
choice of the initial state as $\ket{\text{g}}$, and 
the velocity field around $z=0$ evoles in time as $-A(t)$ according to the
two-state analysis (Fig.~\ref{fig: 1locg}(c)).
Due to the phase-lag between $A(t)$ and $E(t)$, 
the quantum trajectories are navigated from $z=R/2$ to $z=-R/2$
during the $4.5<t<5$, for example, while the electric force $-E(t)$ points 
in the opposite direction during $4.75<t<5$. As a consequence, the 
trajectories are always accumulated at the upper potential
well, in contradiction to our classical intuition.

\begin{figure}[tbhp]
\begin{center}
\end{center}
\includegraphics[width=.9\linewidth]{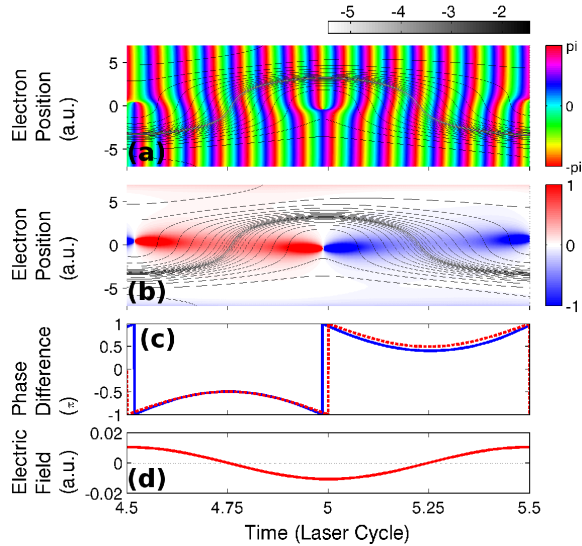}
\caption{Same as Fig.~\ref{fig: 1locg} except that here the system was
  initially prepared in the first excited state $\ket{\text{u}}$.}
\label{fig: 1locu}
\end{figure}

If, instead, the system is initially prepared in the first excited state $\ket{\text{u}}$, 
the initial phase difference $\alpha^{(\text{2lev})}_{\text{LR}}(t_0) = -\pi$.
Due to this offset, the quantum trajectories (and the electron probability density)
are navigated toward the lower potential well in this case. 
In fact, we can see that this is the case for the results presented in Fig.~\ref{fig: 1locu}
where we applied the same laser pulse as used for the results shown in Fig.~\ref{fig: 1locg}
but prepared the system in the 1D model of H$_2^+$ in the first excited state.

\begin{figure}[tbhp]
\begin{center}
\end{center}
\includegraphics[width=.9\linewidth]{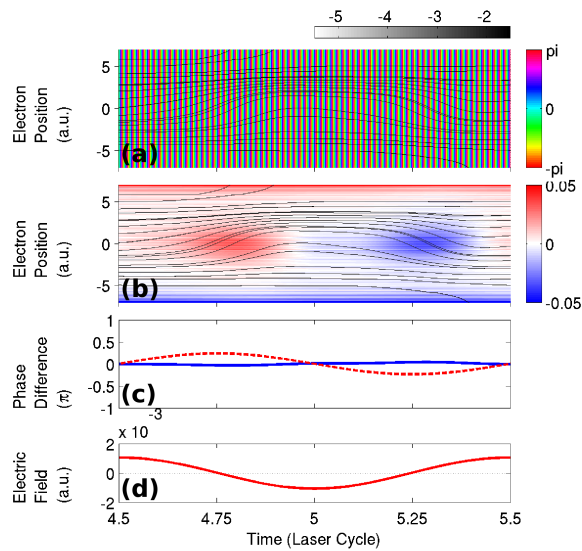}
\caption{Same as Fig.~\ref{fig: 1locg} except that here a laser pulse with 
long wavelength ($5000$ nm) and low intensity ($4\times 10^{10}$ W/cm$^2$) 
(duration $T=10$ cyc, CEP $\phi=0$) was applied.
}
\label{fig: 1loc classical}
\end{figure}

We may finally ask if there is a parameter regime in which 
our intuitive (classical) picture that the electron 
should move in the direction of the electric force is always recovered. 
As we mentioned above, our two-state analysis is based on the assumption 
that $V^{(\text{2lev})}_{\text{L}}$ is the dominant term, i.e.
that the laser-molecule
coupling is strong and/or the photon energy is sufficiently 
larger than the tunnel splitting. 
Please note that the photon energy at the wavelength of $1064$ nm
is $\omega = 0.043$, whereas the tunnel splitting is 
$\Delta_0=0.028$ for the 1D model at $R=7$.
If we decrease the photon energy as well as the laser intensity, the
two-state analysis is no longer applicable. In fact, the results in 
Fig.~\ref{fig: 1loc classical} show that at a wavelength of $5000$ nm and 
an intensity of $4\times 10^{10}$ W/cm$^2$ our 
two-state analysis breaks down, since 
the phase difference $\alpha^{(\text{2lev})}_{\text{LR}}(t)$
(red dashed line) predicted by the two-state analysis deviates qualitatively
from the value of $\alpha_{\text{LR}}(t)$ (blue solid line) calculated 
from the TDSE solution (Fig.~\ref{fig: 1loc classical}(c)).
We can see that in this parameter regime, in which the conventional weak field
perturbation theory may be applied, the motion of the quantum trajectories (and
the electron probability density) follows the electric force of the laser field, and the
intuitive (classical) picture is indeed recovered.

\section{Conclusions}\label{sec: Conclusion}
We have presented an analysis of the attosecond electron dynamics in hydrogen molecular 
ion driven by an intense laser pulse in terms of Bohmian trajectory calculations
using a 1D model. Results of these calculations are found to be in good overall agreement 
with those of numerical simulations of the TDSE.
Recently predicted phenomena such attosecond transient electron localization and 
multiple bursts of ionization within a half cycle of the laser pulse are clearly 
represented by the Bohmian trajectories. Further analysis let us identify the origin 
of the sometimes counter-intuitive motion of the Bohmian trajectories as due to 
the time evolving phase difference of the wavefunction between the two potential wells 
induced by the electric potential of the laser field. We were able to predict the 
time instants at which the trajectories change their directions within a 
simple two-state model. Following our analysis we showed that exposed to an intense laser 
field the electron dynamics 
in the hydrogen molecular ion often does not follow the (classical) force of the 
laser electric field. Our classical expectations are however recovered in the 
perturbative weak-field limit in which the photon energy and intensity of the laser field 
are both sufficiently small.

\section*{Acknowledgments}\label{sec: Acknowledgements}

We acknowledge Professor T. Kato, Dr. A. Pic{\'o}n, and A. Benseny 
for helpful discussions. 
This work was partially supported by the US National Science 
Foundation.


\begin{thebibliography}{42}
\expandafter\ifx\csname natexlab\endcsname\relax\def\natexlab#1{#1}\fi
\expandafter\ifx\csname bibnamefont\endcsname\relax
  \def\bibnamefont#1{#1}\fi
\expandafter\ifx\csname bibfnamefont\endcsname\relax
  \def\bibfnamefont#1{#1}\fi
\expandafter\ifx\csname citenamefont\endcsname\relax
  \def\citenamefont#1{#1}\fi
\expandafter\ifx\csname url\endcsname\relax
  \def\url#1{\texttt{#1}}\fi
\expandafter\ifx\csname urlprefix\endcsname\relax\def\urlprefix{URL }\fi
\providecommand{\bibinfo}[2]{#2}
\providecommand{\eprint}[2][]{\url{#2}}

\bibitem[{\citenamefont{Bohm}(1952{\natexlab{a}})}]{physRev85_166}
\bibinfo{author}{\bibfnamefont{D.}~\bibnamefont{Bohm}}, \bibinfo{journal}{Phys.
  Rev.} \textbf{\bibinfo{volume}{85}}, \bibinfo{pages}{166}
  (\bibinfo{year}{1952}{\natexlab{a}}).

\bibitem[{\citenamefont{Bohm}(1952{\natexlab{b}})}]{physRev85_180}
\bibinfo{author}{\bibfnamefont{D.}~\bibnamefont{Bohm}}, \bibinfo{journal}{Phys.
  Rev.} \textbf{\bibinfo{volume}{85}}, \bibinfo{pages}{180}
  (\bibinfo{year}{1952}{\natexlab{b}}).

\bibitem[{\citenamefont{Holland}(1993)}]{Holland1993}
\bibinfo{author}{\bibfnamefont{P.~R.} \bibnamefont{Holland}},
  \emph{\bibinfo{title}{The Quantum Theory of Motion: An Account of the de
  Broglie-Bohm Causal Interpretation of Quantum Mechanics}}
  (\bibinfo{publisher}{Cambridge University Press}, \bibinfo{year}{1993}).

\bibitem[{\citenamefont{Wyatt}(2005)}]{Wyatt2005}
\bibinfo{author}{\bibfnamefont{R.~E.} \bibnamefont{Wyatt}},
  \emph{\bibinfo{title}{Quantum Dynamics with Trajectories: Introduction to
  Quantum Hydrodynamics}} (\bibinfo{publisher}{Springer},
  \bibinfo{year}{2005}).

\bibitem[{\citenamefont{Kawata et~al.}(1999)\citenamefont{Kawata, Kono, and
  Fujimura}}]{jcp110_11152}
\bibinfo{author}{\bibfnamefont{I.}~\bibnamefont{Kawata}},
  \bibinfo{author}{\bibfnamefont{H.}~\bibnamefont{Kono}}, \bibnamefont{and}
  \bibinfo{author}{\bibfnamefont{Y.}~\bibnamefont{Fujimura}},
  \bibinfo{journal}{J. Chem. Phys.} \textbf{\bibinfo{volume}{110}},
  \bibinfo{pages}{11152} (\bibinfo{year}{1999}).

\bibitem[{\citenamefont{He et~al.}(2008)\citenamefont{He, Becker, and
  Thumm}}]{prl101_213002}
\bibinfo{author}{\bibfnamefont{F.}~\bibnamefont{He}},
  \bibinfo{author}{\bibfnamefont{A.}~\bibnamefont{Becker}}, \bibnamefont{and}
  \bibinfo{author}{\bibfnamefont{U.}~\bibnamefont{Thumm}},
  \bibinfo{journal}{Phys. Rev. Lett.} \textbf{\bibinfo{volume}{101}},
  \bibinfo{pages}{213002} (\bibinfo{year}{2008}).

\bibitem[{\citenamefont{Takemoto and Becker}(2010)}]{prl105_203004}
\bibinfo{author}{\bibfnamefont{N.}~\bibnamefont{Takemoto}} \bibnamefont{and}
  \bibinfo{author}{\bibfnamefont{A.}~\bibnamefont{Becker}},
  \bibinfo{journal}{Phys. Rev. Lett.} \textbf{\bibinfo{volume}{105}},
  \bibinfo{pages}{203004} (\bibinfo{year}{2010}).

\bibitem[{\citenamefont{Zuo and Bandrauk}(1995)}]{pra52_r2511}
\bibinfo{author}{\bibfnamefont{T.}~\bibnamefont{Zuo}} \bibnamefont{and}
  \bibinfo{author}{\bibfnamefont{A.~D.} \bibnamefont{Bandrauk}},
  \bibinfo{journal}{Phys. Rev. A} \textbf{\bibinfo{volume}{52}},
  \bibinfo{pages}{R2511} (\bibinfo{year}{1995}).

\bibitem[{\citenamefont{Mulliken}(1939)}]{jcp7_20}
\bibinfo{author}{\bibfnamefont{R.~S.} \bibnamefont{Mulliken}},
  \bibinfo{journal}{J. Chem. Phys.} \textbf{\bibinfo{volume}{7}},
  \bibinfo{pages}{20} (\bibinfo{year}{1939}).

\bibitem[{\citenamefont{Hu et~al.}(2009)\citenamefont{Hu, Collins, and
  Schneider}}]{pra80_023426}
\bibinfo{author}{\bibfnamefont{S.~X.} \bibnamefont{Hu}},
  \bibinfo{author}{\bibfnamefont{L.~A.} \bibnamefont{Collins}},
  \bibnamefont{and} \bibinfo{author}{\bibfnamefont{B.~I.}
  \bibnamefont{Schneider}}, \bibinfo{journal}{Phys. Rev. A}
  \textbf{\bibinfo{volume}{80}}, \bibinfo{pages}{023426}
  (\bibinfo{year}{2009}).

\bibitem[{\citenamefont{Chelkowski et~al.}(1995)\citenamefont{Chelkowski,
  T.~Zuo, and Bandrauk}}]{pra52_2977}
\bibinfo{author}{\bibfnamefont{S.}~\bibnamefont{Chelkowski}},
  \bibinfo{author}{\bibfnamefont{O.~A.} \bibnamefont{T.~Zuo}},
  \bibnamefont{and} \bibinfo{author}{\bibfnamefont{A.~D.}
  \bibnamefont{Bandrauk}}, \bibinfo{journal}{Phys. Rev. A}
  \textbf{\bibinfo{volume}{52}}, \bibinfo{pages}{2977} (\bibinfo{year}{1995}).

\bibitem[{\citenamefont{Roudnev et~al.}(2004)\citenamefont{Roudnev, Esry, and
  Ben-Itzhak}}]{prl93_163601}
\bibinfo{author}{\bibfnamefont{V.}~\bibnamefont{Roudnev}},
  \bibinfo{author}{\bibfnamefont{B.~D.} \bibnamefont{Esry}}, \bibnamefont{and}
  \bibinfo{author}{\bibfnamefont{I.}~\bibnamefont{Ben-Itzhak}},
  \bibinfo{journal}{Phys. Rev. Lett.} \textbf{\bibinfo{volume}{93}},
  \bibinfo{pages}{163601} (\bibinfo{year}{2004}).

\bibitem[{\citenamefont{He et~al.}(2007)\citenamefont{He, Ruiz, and
  Becker}}]{prl99_083002}
\bibinfo{author}{\bibfnamefont{F.}~\bibnamefont{He}},
  \bibinfo{author}{\bibfnamefont{C.}~\bibnamefont{Ruiz}}, \bibnamefont{and}
  \bibinfo{author}{\bibfnamefont{A.}~\bibnamefont{Becker}},
  \bibinfo{journal}{Phys. Rev. Lett.} \textbf{\bibinfo{volume}{99}},
  \bibinfo{pages}{083002} (\bibinfo{year}{2007}).

\bibitem[{\citenamefont{Takemoto and Becker}()}]{takemoto_tobepublished}
\bibinfo{author}{\bibfnamefont{N.}~\bibnamefont{Takemoto}} \bibnamefont{and}
  \bibinfo{author}{\bibfnamefont{A.}~\bibnamefont{Becker}}, \bibinfo{note}{to
  be published}.

\bibitem[{\citenamefont{Javanainen et~al.}(1988)\citenamefont{Javanainen,
  Eberly, and Su}}]{physRevA38_3430}
\bibinfo{author}{\bibfnamefont{J.}~\bibnamefont{Javanainen}},
  \bibinfo{author}{\bibfnamefont{J.~H.} \bibnamefont{Eberly}},
  \bibnamefont{and} \bibinfo{author}{\bibfnamefont{Q.}~\bibnamefont{Su}},
  \bibinfo{journal}{Phys. Rev. A} \textbf{\bibinfo{volume}{38}},
  \bibinfo{pages}{3430} (\bibinfo{year}{1988}).

\bibitem[{\citenamefont{Kulander et~al.}(1996)\citenamefont{Kulander, Mies, and
  Schafer}}]{physRevA53_2562}
\bibinfo{author}{\bibfnamefont{K.~C.} \bibnamefont{Kulander}},
  \bibinfo{author}{\bibfnamefont{F.~H.} \bibnamefont{Mies}}, \bibnamefont{and}
  \bibinfo{author}{\bibfnamefont{K.~J.} \bibnamefont{Schafer}},
  \bibinfo{journal}{Phys. Rev. A} \textbf{\bibinfo{volume}{53}},
  \bibinfo{pages}{2562} (\bibinfo{year}{1996}).

\bibitem[{\citenamefont{Marston and Balint-Kurti}(1989)}]{jcp91_3571}
\bibinfo{author}{\bibfnamefont{C.~C.} \bibnamefont{Marston}} \bibnamefont{and}
  \bibinfo{author}{\bibfnamefont{G.~G.} \bibnamefont{Balint-Kurti}},
  \bibinfo{journal}{J. Chem. Phys.} \textbf{\bibinfo{volume}{91}},
  \bibinfo{pages}{3571} (\bibinfo{year}{1989}).

\bibitem[{\citenamefont{Bandrauk and Shen}(1993)}]{jcp99_1185}
\bibinfo{author}{\bibfnamefont{A.~D.} \bibnamefont{Bandrauk}} \bibnamefont{and}
  \bibinfo{author}{\bibfnamefont{H.}~\bibnamefont{Shen}}, \bibinfo{journal}{J.
  Chem. Phys.} \textbf{\bibinfo{volume}{99}}, \bibinfo{pages}{1185}
  (\bibinfo{year}{1993}).

\bibitem[{\citenamefont{Takahashi and Ikeda}(1993)}]{jcp99_8680}
\bibinfo{author}{\bibfnamefont{K.}~\bibnamefont{Takahashi}} \bibnamefont{and}
  \bibinfo{author}{\bibfnamefont{K.}~\bibnamefont{Ikeda}}, \bibinfo{journal}{J.
  Chem. Phys.} \textbf{\bibinfo{volume}{99}}, \bibinfo{pages}{8680}
  (\bibinfo{year}{1993}).

\bibitem[{\citenamefont{Press et~al.}(1992)\citenamefont{Press, Teukolsky,
  Vetterling, and Flannery}}]{Press_etal1992}
\bibinfo{author}{\bibfnamefont{W.~H.} \bibnamefont{Press}},
  \bibinfo{author}{\bibfnamefont{S.~A.} \bibnamefont{Teukolsky}},
  \bibinfo{author}{\bibfnamefont{W.~T.} \bibnamefont{Vetterling}},
  \bibnamefont{and} \bibinfo{author}{\bibfnamefont{B.~P.}
  \bibnamefont{Flannery}}, \emph{\bibinfo{title}{Numerical Recipes in Fortran
  77: The Art of Scientific Computing, Vol. 1 of Fortran Numerical Recipes}}
  (\bibinfo{publisher}{Cambridge University Press}, \bibinfo{address}{New
  York}, \bibinfo{year}{1992}), \bibinfo{edition}{2nd} ed.

\bibitem[{\citenamefont{Garashchuk and Rassolov}(2002)}]{chemPhysLett364_562}
\bibinfo{author}{\bibfnamefont{S.}~\bibnamefont{Garashchuk}} \bibnamefont{and}
  \bibinfo{author}{\bibfnamefont{V.~A.} \bibnamefont{Rassolov}},
  \bibinfo{journal}{Chem. Phys. Lett.} \textbf{\bibinfo{volume}{364}},
  \bibinfo{pages}{562} (\bibinfo{year}{2002}).

\bibitem[{\citenamefont{Garashchuk and Rassolov}(2003)}]{jcp118_2482}
\bibinfo{author}{\bibfnamefont{S.}~\bibnamefont{Garashchuk}} \bibnamefont{and}
  \bibinfo{author}{\bibfnamefont{V.~A.} \bibnamefont{Rassolov}},
  \bibinfo{journal}{J. Chem. Phys.} \textbf{\bibinfo{volume}{118}},
  \bibinfo{pages}{2482} (\bibinfo{year}{2003}).

\bibitem[{\citenamefont{Lopreore and Wyatt}(1999)}]{prl82_5190}
\bibinfo{author}{\bibfnamefont{C.~L.} \bibnamefont{Lopreore}} \bibnamefont{and}
  \bibinfo{author}{\bibfnamefont{R.~E.} \bibnamefont{Wyatt}},
  \bibinfo{journal}{Phys. Rev. Lett.} \textbf{\bibinfo{volume}{82}},
  \bibinfo{pages}{5190} (\bibinfo{year}{1999}).

\bibitem[{\citenamefont{Mayor et~al.}(1999)\citenamefont{Mayor, Askar, and
  Rabitz}}]{jcp111_2423}
\bibinfo{author}{\bibfnamefont{F.~S.} \bibnamefont{Mayor}},
  \bibinfo{author}{\bibfnamefont{A.}~\bibnamefont{Askar}}, \bibnamefont{and}
  \bibinfo{author}{\bibfnamefont{H.~A.} \bibnamefont{Rabitz}},
  \bibinfo{journal}{J. Chem. Phys.} \textbf{\bibinfo{volume}{111}},
  \bibinfo{pages}{2423} (\bibinfo{year}{1999}).

\bibitem[{\citenamefont{Poirier}(2010)}]{chemPhys370_4}
\bibinfo{author}{\bibfnamefont{B.}~\bibnamefont{Poirier}},
  \bibinfo{journal}{Chem. Phys.} \textbf{\bibinfo{volume}{370}},
  \bibinfo{pages}{4} (\bibinfo{year}{2010}).

\bibitem[{\citenamefont{Botheron and Pons}(2010)}]{pra82_021404}
\bibinfo{author}{\bibfnamefont{P.}~\bibnamefont{Botheron}} \bibnamefont{and}
  \bibinfo{author}{\bibfnamefont{B.}~\bibnamefont{Pons}},
  \bibinfo{journal}{Phys. Rev. A} \textbf{\bibinfo{volume}{82}},
  \bibinfo{pages}{021404(R)} (\bibinfo{year}{2010}).

\bibitem[{\citenamefont{Sharp}(1970)}]{atDataNucDataTables2_119}
\bibinfo{author}{\bibfnamefont{T.}~\bibnamefont{Sharp}}, \bibinfo{journal}{At.
  Data Nuc. Data Tables} \textbf{\bibinfo{volume}{2}}, \bibinfo{pages}{119}
  (\bibinfo{year}{1970}).

\bibitem[{\citenamefont{Faisal and Schwengelbeck}(1998)}]{pramana51_585}
\bibinfo{author}{\bibfnamefont{F.}~\bibnamefont{Faisal}} \bibnamefont{and}
  \bibinfo{author}{\bibfnamefont{U.}~\bibnamefont{Schwengelbeck}},
  \bibinfo{journal}{Pramana} \textbf{\bibinfo{volume}{51}},
  \bibinfo{pages}{585} (\bibinfo{year}{1998}).

\bibitem[{\citenamefont{Lai et~al.}(2009{\natexlab{a}})\citenamefont{Lai, Cai,
  and Zhan}}]{euroPhysJD53_393}
\bibinfo{author}{\bibfnamefont{X.~Y.} \bibnamefont{Lai}},
  \bibinfo{author}{\bibfnamefont{Q.~Y.} \bibnamefont{Cai}}, \bibnamefont{and}
  \bibinfo{author}{\bibfnamefont{M.~S.} \bibnamefont{Zhan}},
  \bibinfo{journal}{Eur. Phys. J. D} \textbf{\bibinfo{volume}{53}},
  \bibinfo{pages}{393} (\bibinfo{year}{2009}{\natexlab{a}}).

\bibitem[{\citenamefont{Lai et~al.}(2009{\natexlab{b}})\citenamefont{Lai, Cai,
  and Zhan}}]{newJPhys11_113035}
\bibinfo{author}{\bibfnamefont{X.~Y.} \bibnamefont{Lai}},
  \bibinfo{author}{\bibfnamefont{Q.-Y.} \bibnamefont{Cai}}, \bibnamefont{and}
  \bibinfo{author}{\bibfnamefont{M.~S.} \bibnamefont{Zhan}},
  \bibinfo{journal}{New J. Phys.} \textbf{\bibinfo{volume}{11}},
  \bibinfo{pages}{113035} (\bibinfo{year}{2009}{\natexlab{b}}).

\bibitem[{\citenamefont{Burnett et~al.}(1998)\citenamefont{Burnett, Watson,
  Sanpera, and Knight}}]{philTransRoySocLondonA356_317}
\bibinfo{author}{\bibfnamefont{K.}~\bibnamefont{Burnett}},
  \bibinfo{author}{\bibfnamefont{J.~B.} \bibnamefont{Watson}},
  \bibinfo{author}{\bibfnamefont{A.}~\bibnamefont{Sanpera}}, \bibnamefont{and}
  \bibinfo{author}{\bibfnamefont{P.~L.} \bibnamefont{Knight}},
  \bibinfo{journal}{Phil. Trans. Roy. Soc. London A}
  \textbf{\bibinfo{volume}{356}}, \bibinfo{pages}{317} (\bibinfo{year}{1998}).

\bibitem[{\citenamefont{D{\"{o}}rr}(2000)}]{optExpress6_111}
\bibinfo{author}{\bibfnamefont{M.}~\bibnamefont{D{\"{o}}rr}},
  \bibinfo{journal}{Opt. Express} \textbf{\bibinfo{volume}{6}},
  \bibinfo{pages}{111} (\bibinfo{year}{2000}).

\bibitem[{\citenamefont{Baier et~al.}(2006)\citenamefont{Baier, Ruiz, Plaja,
  and Becker}}]{physRevA74_033405}
\bibinfo{author}{\bibfnamefont{S.}~\bibnamefont{Baier}},
  \bibinfo{author}{\bibfnamefont{C.}~\bibnamefont{Ruiz}},
  \bibinfo{author}{\bibfnamefont{L.}~\bibnamefont{Plaja}}, \bibnamefont{and}
  \bibinfo{author}{\bibfnamefont{A.}~\bibnamefont{Becker}},
  \bibinfo{journal}{Phys. Rev. A} \textbf{\bibinfo{volume}{74}},
  \bibinfo{pages}{033405} (\bibinfo{year}{2006}).

\bibitem[{\citenamefont{Kayanuma}(1993)}]{physRevB47_9940}
\bibinfo{author}{\bibfnamefont{Y.}~\bibnamefont{Kayanuma}},
  \bibinfo{journal}{Phys. Rev. B} \textbf{\bibinfo{volume}{47}},
  \bibinfo{pages}{9940} (\bibinfo{year}{1993}).

\bibitem[{\citenamefont{Kayanuma}(1994)}]{physRevA50_843}
\bibinfo{author}{\bibfnamefont{Y.}~\bibnamefont{Kayanuma}},
  \bibinfo{journal}{Phys. Rev. A} \textbf{\bibinfo{volume}{50}},
  \bibinfo{pages}{843} (\bibinfo{year}{1994}).

\bibitem[{\citenamefont{Creffield}(2003)}]{physRevB67_165301}
\bibinfo{author}{\bibfnamefont{C.}~\bibnamefont{Creffield}},
  \bibinfo{journal}{Phys. Rev. B} \textbf{\bibinfo{volume}{67}},
  \bibinfo{pages}{165301} (\bibinfo{year}{2003}).

\bibitem[{\citenamefont{Grossmann et~al.}(1991)\citenamefont{Grossmann,
  Dittrich, Jung, and H\"anggi}}]{prl67_516}
\bibinfo{author}{\bibfnamefont{F.}~\bibnamefont{Grossmann}},
  \bibinfo{author}{\bibfnamefont{T.}~\bibnamefont{Dittrich}},
  \bibinfo{author}{\bibfnamefont{P.}~\bibnamefont{Jung}}, \bibnamefont{and}
  \bibinfo{author}{\bibfnamefont{P.}~\bibnamefont{H\"anggi}},
  \bibinfo{journal}{Phys. Rev. Lett.} \textbf{\bibinfo{volume}{67}},
  \bibinfo{pages}{516} (\bibinfo{year}{1991}).

\bibitem[{\citenamefont{Gro{\ss}mann and
  H{\"{a}}nggi}(1992)}]{europhysLett18_571}
\bibinfo{author}{\bibfnamefont{F.}~\bibnamefont{Gro{\ss}mann}}
  \bibnamefont{and}
  \bibinfo{author}{\bibfnamefont{P.}~\bibnamefont{H{\"{a}}nggi}},
  \bibinfo{journal}{Europhys. Lett.} \textbf{\bibinfo{volume}{18}},
  \bibinfo{pages}{571} (\bibinfo{year}{1992}).

\bibitem[{\citenamefont{Yu.{ }Ivanov and Corkum}(1993)}]{physRevA48_580}
\bibinfo{author}{\bibfnamefont{M.}~\bibnamefont{Yu.{ }Ivanov}}
  \bibnamefont{and} \bibinfo{author}{\bibfnamefont{P.~B.}
  \bibnamefont{Corkum}}, \bibinfo{journal}{Phys. Rev. A}
  \textbf{\bibinfo{volume}{48}}, \bibinfo{pages}{580} (\bibinfo{year}{1993}).

\bibitem[{\citenamefont{Yu.{ }Ivanov et~al.}(1993)\citenamefont{Yu.{ }Ivanov,
  Corkum, and Dietrich}}]{laserPhys3_375}
\bibinfo{author}{\bibfnamefont{M.}~\bibnamefont{Yu.{ }Ivanov}},
  \bibinfo{author}{\bibfnamefont{P.~B.} \bibnamefont{Corkum}},
  \bibnamefont{and} \bibinfo{author}{\bibfnamefont{P.}~\bibnamefont{Dietrich}},
  \bibinfo{journal}{Laser Phys.} \textbf{\bibinfo{volume}{3}},
  \bibinfo{pages}{375} (\bibinfo{year}{1993}).

\bibitem[{\citenamefont{Burdov}(1997)}]{JETP85_657}
\bibinfo{author}{\bibfnamefont{V.~A.} \bibnamefont{Burdov}},
  \bibinfo{journal}{JETP} \textbf{\bibinfo{volume}{85}}, \bibinfo{pages}{657}
  (\bibinfo{year}{1997}).

\bibitem[{\citenamefont{Santana et~al.}(2001)\citenamefont{Santana, Gomez{
  }Llorente, and Delgado}}]{jPhysB34_2371}
\bibinfo{author}{\bibfnamefont{A.}~\bibnamefont{Santana}},
  \bibinfo{author}{\bibfnamefont{J.~M.} \bibnamefont{Gomez{ }Llorente}},
  \bibnamefont{and} \bibinfo{author}{\bibfnamefont{V.}~\bibnamefont{Delgado}},
  \bibinfo{journal}{J. Phys. B} \textbf{\bibinfo{volume}{34}},
  \bibinfo{pages}{2371} (\bibinfo{year}{2001}).

\end{thebibliography}
%

\end{document}